\documentclass{iaus}
\usepackage{graphicx}

\title[Stellar ages from asteroseismology] 
{Stellar ages from asteroseismology}

\author[Yveline Lebreton \& Josefina Montalb{\'a}n]   %% give here short author list %%
{Yveline Lebreton$^{1, 2}$
 \and Josefina Montalb{\'a}n$^3$}

\affiliation{$^1$ Observatoire de Paris, GEPI, UMR CNRS 8111, \\ Place J. Janssen,
92195 Meudon, France \\ email: {\tt Yveline.Lebreton@obspm.fr} \\[\affilskip]
$^2$ IPR, Universit{\'e} de Rennes 1, 35042 Rennes, France\\[\affilskip]
$^3$ Institut d'Astrophysique et de G\'{e}ophysique, Universit\'{e} de 
Li\`{e}ge, Belgium\\email: {\tt j.montalban@ulg.ac.be}}

\pubyear{2008}
\volume{258}  %% insert here IAU Symposium No.
\pagerange{119--126}
\jname{The Ages of Stars}
\editors{A.C. Editor, B.D. Editor \& C.E. Editor, eds.}
\begin{document}

\maketitle

\begin{abstract}
Asteroseismology has been recognized for a long time as a very powerful mean to probe stellar interiors. The oscillations frequencies are closely related to stellar internal structure properties via the density and the sound speed profiles. Since these properties are in turn tightly linked with the mass and evolutionary state, we can expect to determine the age and mass of a star from the comparison of its oscillation spectrum with the predictions of stellar models. Such a comparison will of course suffer both from the problems  we face when modeling a particular star (for instance the uncertainties on its global parameters and chemical composition) and from our general misunderstanding of the physical processes at work in stellar interiors (for instance the various transport processes that may lead to core mixing and affect the ages predicted by models). However for stars where observations have provided very precise and numerous oscillation frequencies together with accurate global parameters and additional information (as the radius or the mass of the star if it is member of a binary system, the radius if it observable in interferometry or the mean density if the star is an exoplanet host), we can also expect to better constrain the physical description of the stellar structure and transport processes and to finally get a more reliable age estimation.

After a brief survey of stellar pulsations, we present some general seismic diagnostics that can be used to infer the age of a pulsating star as well as their limitations. We then illustrate the ability of asteroseismology to scrutinize stellar interiors on the basis of a few exemples. In the years to come, extended very precise asteroseismic observations are expected, either in photometry or in spectroscopy, from present and future ground-based ({\small{HARPS, CORALIE, ELODIE, UVES, UCLES, SIAMOIS, SONG}}) or spatial devices ({\small{MOST, CoRoT, WIRE, Kepler, PLATO}}). This will considerably enlarge the sample of stars eligible to asteroseismic age determination and should allow to estimate the age of individual stars with a 10-20\% accuracy.

\keywords{stars: oscillations, stars: interiors, stars: evolution, stars: fundamental parameters}
%% add here a maximum of 10 keywords, to be taken form the file <Keywords.txt>
\end{abstract}

\firstsection % if your document starts with a section,
              % remove some space above using this command.
\section{Introduction}

\begin{figure}[t]
% \vspace*{-2.0 cm}
\begin{center}
\resizebox*{0.45\hsize}{!}{\includegraphics*{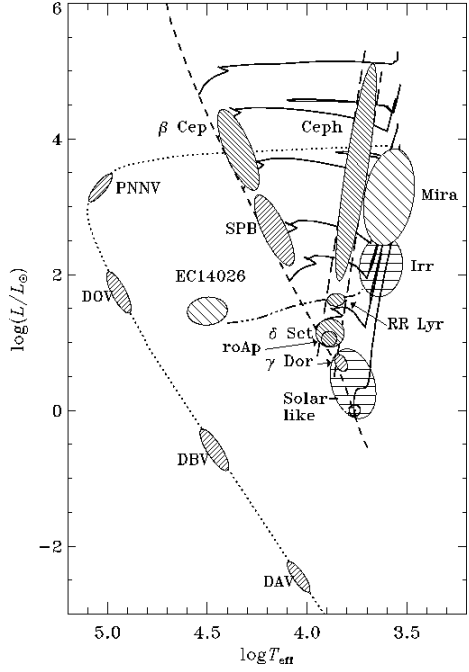}}
% \includegraphics[width=3.4in]{fig1.eps} 
% \vspace*{-1.0 cm}
 \caption{Pulsators in the HR diagram (courtesy J. Christensen-Dalsgaard).}
   \label{fig1}
\end{center}
\end{figure}

As can be seen in Fig.\,\ref{fig1}, pulsating stars are presently observed in nearly each region of the HR diagram where stars are observed. Stellar pulsations may be excited by different mechanisms, in a large range of amplitudes. Self-excitation results from the $\kappa$-mechanism which drives pulsations either in the HeII ionisation zone (oscillations in $\delta$ Scuti, Cepheids, RR Lyrae or DB white dwarfs), in the HI and HeI ionisation zones  (oscillations in roAp, Miras and irregular variables) or in the metal opacity bump (see e.g,  \cite[Dziembowski et al. 1993]{Dziemetal93} ; \cite[Pamiatnykh 1999]{Pamyat99} for oscillations in $\beta$ Cephei and slowly pulsating -SPB- stars). In $\gamma$ Doradus stars, excitations are interpreted as due to the convective blocking of the radiative flux (see e.g, \cite[Guzik et al. 2000]{Guzikal00} ; \cite[Warner et al. 2003]{Warneetal03} ; \cite[Dupret et al. 2004]{Dupreetal04}). Finally, in stars with significant convective envelopes, the so-called solar-like oscillations are interpreted as resulting from stochastic excitation by turbulent convective motions. This can occur on the main sequence (\cite[Christensen-Dalsgaard 1982]{JCD82}) and in subgiants and red giants (\cite[Dziembowski et al. 2001]{Dziemetal01}).

Oscillations are observed either by spectroscopic measurements (velocity variations) or by photometric ones (intensity variations). While only one or two modes but with large amplitudes are observed in Cepheids, W Virginis or RR Lyrae stars, a large number of modes with small amplitudes are observable in solar-like pulsators, $\delta$ Scuti, roAp, $\beta$ Cephei, SPB, $\gamma$ Doradus and white dwarfs pulsators. In solar-like pulsators the variations in velocity amplitudes range from less than $1.0$ to about $50\ \rm{m.s}^{-1}$ (Sun: $0.2\ \rm{m.s}^{-1}$) while relative intensity changes vary from $10^{-6}$ to $10^{-3}$ (Sun: $4.\ 10^{-6}$) and the amplitude scales as $\left({\frac{L}{M}}\right)^{\alpha}$ where $\alpha$ is in the range $0.7-1.0$ (\cite[Kjeldsen and Bedding 1995]{Kjeldsen95} ; \cite[Samadi et al. 2005]{Samadetal05}). Amplitudes in $\delta$ Scuti, roAp, $\beta$ Cephei, SPB, $\gamma$ Doradus and white dwarfs pulsators range from $10^{-3}$ to a few $10^{-1}$ mag.

\section{Characteristics of stellar oscillations}
\label{section2}
Each oscillation eigenmode of frequency $\nu_{n, \ell, m}$ consists in a radial part characterized by the radial order $n$ ($n$ is the number of nodes along the stellar radius) and in a surface pattern characterized by a spherical harmonic $Y_{\ell, m}$ where $\ell$ is the mode angular degree ($\ell$ is a measure of the number of wavelengths along the stellar circumference) and $m$ is the azimutal order (number of nodes along the equator).

In the resolution of the equations of stellar oscillations two characteristic frequencies appear, in addition to the cut-off frequency. First, the acoustic (Lamb) frequency 
%$S_\ell=[\ell(\ell+1)]^\frac{1}{2} \frac{c}{r}$ 
$S_\ell=[\ell(\ell+1)]^\frac{1}{2} ({c}/{r})$ 
which is a measure of the compressibility of the medium and depends on the adiabatic sound speed 
%$c=\left({\frac{\Gamma_1 P}{\rho}}\right)^\frac{1}{2}$ 
$c=({{\Gamma_1 P}/{\rho}})^\frac{1}{2}$ 
where 
$\Gamma_1=({\partial \ln P}/{\partial\ln \rho})_{\rm {ad}}$ 
%$\Gamma_1=\left(\frac{\partial \ln P}{\partial\ln \rho}\right)_{\rm {ad}}$ 
is the first adiabatic coefficient. Second, the buoyancy or Brunt-V$\ddot{\rm a}$is$\ddot{\rm a}$l$\ddot{\rm a}$ frequency $N_{\rm BV}$, such that $N^2_{\rm BV} = g\left( \frac{1}{\Gamma_1}\frac{d\ln P}{dr}- \frac{d\ln \rho}{dr}\right)$. $N^2_{\rm BV}$ is positive in a radiative region where it corresponds to the oscillation frequency of a perturbed fluid element while it is negative in convective regions. 
It is worth noticing that for ideal gases, 
%$c\propto \left({\frac{T}{\mu}}\right)^\frac{1}{2}$ 
$c\propto ({{T}/{\mu}})^\frac{1}{2}$ 
and $N^2_{\rm BV} \simeq ({g^2 \rho}/{P}) (\nabla_{\rm {ad}}-\nabla+\nabla_{\mu})$ where $\mu$ is the mean molecular weight and $\nabla_{\mu}$ is its gradient.

It can be shown that different waves can propagate inside a star: modes with frequencies higher than both $S_\ell$ and $N_{\rm BV}$ correspond to standing sound waves (acoustic pressure modes or p-modes) while modes with frequencies lower than both $S_\ell$ and $N_{\rm BV}$ correspond to standing gravity waves (g-modes). Otherwise modes are evanescent. In absence of rotation, the frequency of an oscillation mode only depends on two parameters (density $\rho$ and $\Gamma_1$ or equivalently $\rho$ and sound speed). Therefore frequencies change with mass, evolution (age) and chemical composition.

\begin{figure}[t]
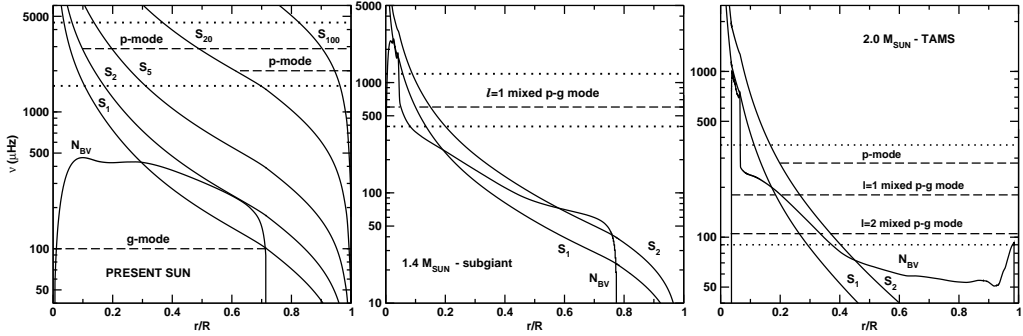

% \vspace*{-2.0 cm}
\begin{center}
\resizebox*{\hsize}{!}{\includegraphics*{fig2a.eps}\includegraphics*{fig2b.eps}
\includegraphics*{fig2c.eps}}
% \resizebox{\includegraphics[width=3.4in]{fig2a.eps} }
% \vspace*{-1.0 cm}
 \caption{Propagation diagrams for a solar model (left), a $1.4\ M_\odot$ model on the subgiant branch (centre) and a $2\ M_\odot$ model ($\delta$ Scuti) at the end of the MS (right). The frequency domains of solar-like oscillations (high degree p-modes) is located in-between the dotted lines.}
   \label{fig2}
\end{center}
\end{figure}

In Fig.~\ref{fig2}, we show propagation diagrams. The left figure shows that in the present solar interior the domains of propagation of p and g-modes are well separated. The central figure shows a model of a star of $1.4\ M_\odot$ on the subgiant branch. In this star a $\mu$-gradient has been built in the radiative core during the main sequence (MS) evolution and the core has become more and more dense on the subgiant branch. As a result, the Brunt-V$\ddot{\rm a}$is$\ddot{\rm a}$l$\ddot{\rm a}$ frequency shows a central peak; this allows g-modes to propagate in a domain of frequency corresponding to the p-modes domain. Due to the closeness of the propagation regions (see the $\ell=1$ mode on the figure), a g-mode can interact with a p-mode of the same frequency through the so-called avoided crossing and we therefore expect to observe p--g mixed modes in this kind of stars (\cite[Aizenman et al. 1977]{1977AA5841A}). This situation is also expected to happen in $\delta$ Scuti stars (right figure), on the MS, because of the building of a $\mu$-gradient in the radiative regions just above the receding convective core. More generally, depending on the mass and evolutionary state, different kinds of modes are predicted to be observable: for instance in $\beta$ Cephei and $\delta$ Scuti stars low order p and g modes are expected with periods in the range 2--8 hours for the former and 30 min--6 hours for the latter, in SPB stars and $\gamma$ Doradus stars high order g-modes are expected with periods in the range 15 hours--5 days for the former and 8 hours--3 days for the latter while in solar-like pulsators high order p-modes are expected with periods in the range a few minutes (MS stars) to a few hours (red giants).

\section{Age diagnostics through p-modes}

In the case of the Sun, more than $10^5$ modes can be observed, of all degrees, with an excellent precision. It is therefore possible to derive the density, the sound speed or the internal rotation rate by inversion of the frequencies which in turn gives access to very precise information on the solar properties such as the age, helium content and depth of the outer convective envelope (see e.g. Christensen-Dalsgaard, these proceedings). On the other hand, in the case of solar-like pulsators, only a few modes of low degree (mainly $\ell= 0, 1, 2, 3$) are accessible to observations. However, as discussed below, this can provide valuable information on the stellar interior.

\begin{figure}[t]
% \vspace*{-2.0 cm}
\begin{center}
\resizebox*{0.7\hsize}{!}{\includegraphics*{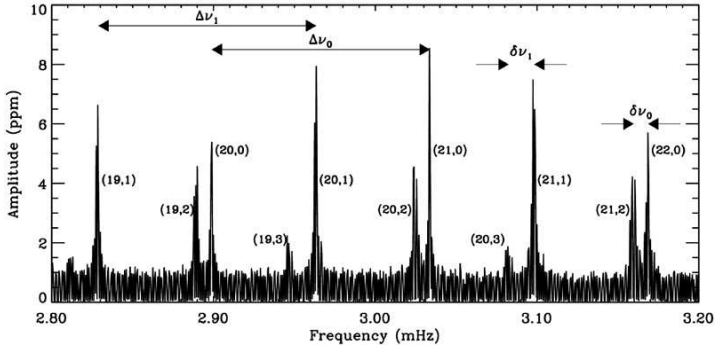}}
%\resizebox*{\includegraphics*[width=3.4in]{fig3.eps}}
% \vspace*{-1.0 cm}
 \caption{Detail of the solar oscillations amplitude spectrum, as observed by VIRGO on SOHO (Figure from \cite[Bedding \& Kjeldsen 2003]{2003PASA...20..203B}). Each frequency peak is labelled with the value of the mode order and degree and the large and small separations are indicated.}
\label{fig3}
\end{center}
\end{figure}

\begin{figure}[b]
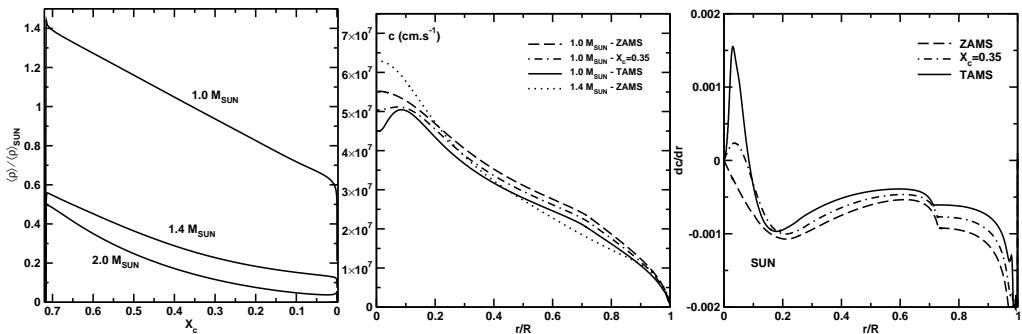

% \vspace*{-2.0 cm}
\begin{center}
\resizebox*{\hsize}{!}{\includegraphics*{fig4a.eps}\includegraphics*{fig4b.eps}\includegraphics*{fig4c.eps}}
% \resizebox{\includegraphics[width=3.4in]{fig2a.eps} }
% \vspace*{-1.0 cm}
 \caption{Left: the change of the mean density with evolution on the MS for 3 values of the stellar mass. Centre and right: the sound speed profile (centre) and the $dc/dr$ profile (right) in $1\ M_\odot$ models, on the ZAMS, in the middle of the MS and at the end of the MS (TAMS)}
   \label{fig4}
\end{center}
\end{figure}

\subsection{Asymptotic theory: frequency combinations and asteroseismic diagrams}

Figure~\ref{fig3} shows a detail of the solar power spectrum where the modes have been identified by their order and degree. The regular pattern seen is indeed predicted by the asymptotic theory of stellar oscillations (\cite[Vandakurov 1967]{1967AZh....44..786V}, \cite[Tassoul 1980]{1980ApJS...43..469T}) from which the frequency of a mode of high order $n$ and of degree $\ell\ll n$ is given by $\nu_{\rm n,\ell}= \Delta \nu\left( n+\frac{1}{2}\ell +\epsilon_{n, \ell}\right)-\ell(\ell+1) D_0$ where $\epsilon_{n, \ell}$ is a function sensitive to surface physics, but only weakly sensitive to the order and degree of the mode, and $\Delta\nu$ and $D_0$ will be explicited below. As a consequence the difference in frequency between two modes of consecutive orders and same degree is approximately constant and given by $\Delta \nu\simeq \nu_{\rm n+1,\ell}-\nu_{\rm n,\ell}\equiv\Delta \nu_{\ell}$ while the difference in frequency between two modes of consecutive orders and degrees differing by two units is $\delta \nu_{\ell}\equiv \nu_{\rm n,\ell}-\nu_{\rm n-1,\ell+2}=4(\ell+6)D_0$.  The differences $\Delta \nu$ and $\delta \nu$ are called the large and small frequency separations, respectively.

\begin{figure}[b]
% \vspace*{-2.0 cm}
\begin{center}
\resizebox*{\hsize}{!}{\includegraphics*{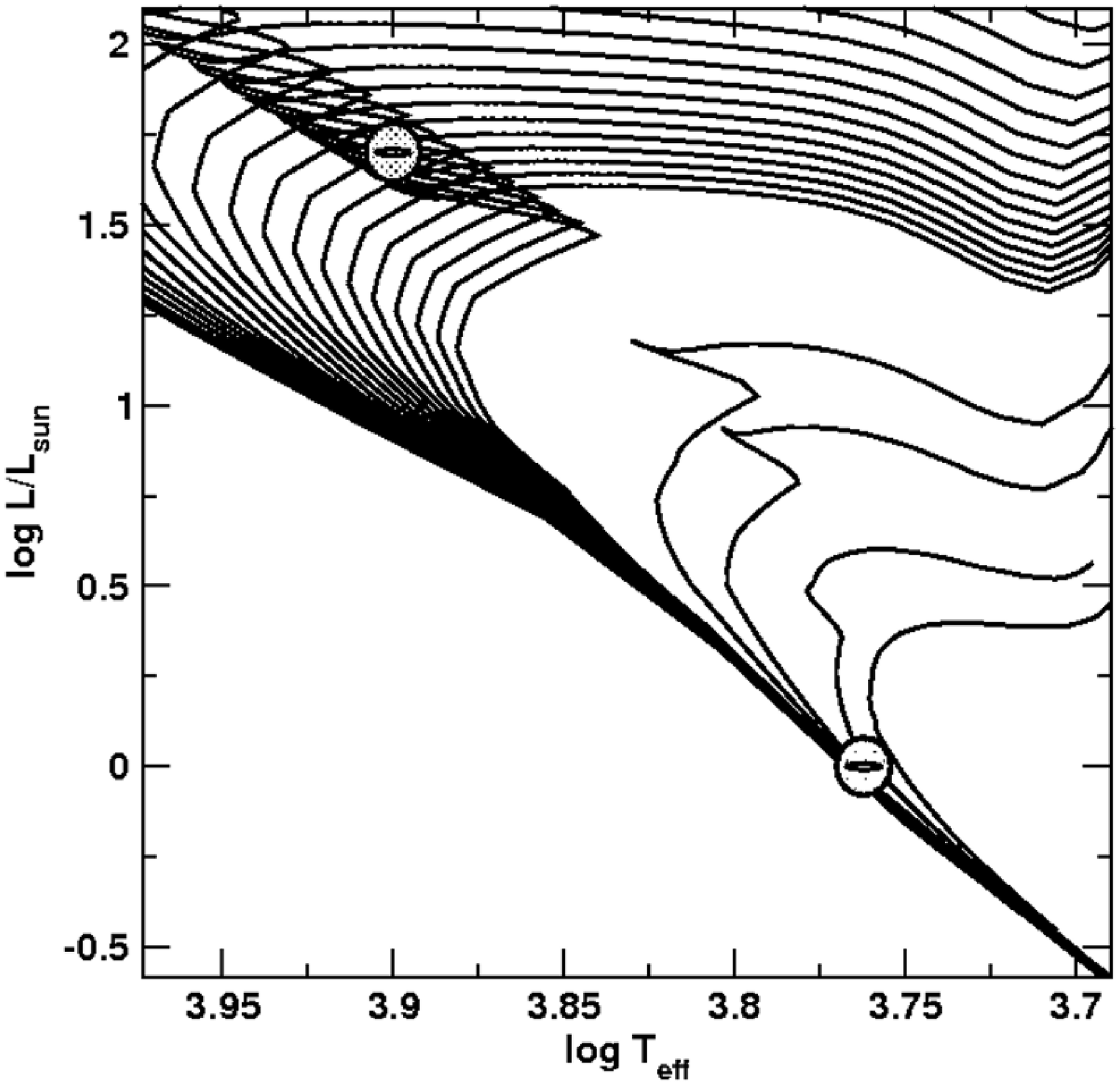}\includegraphics*{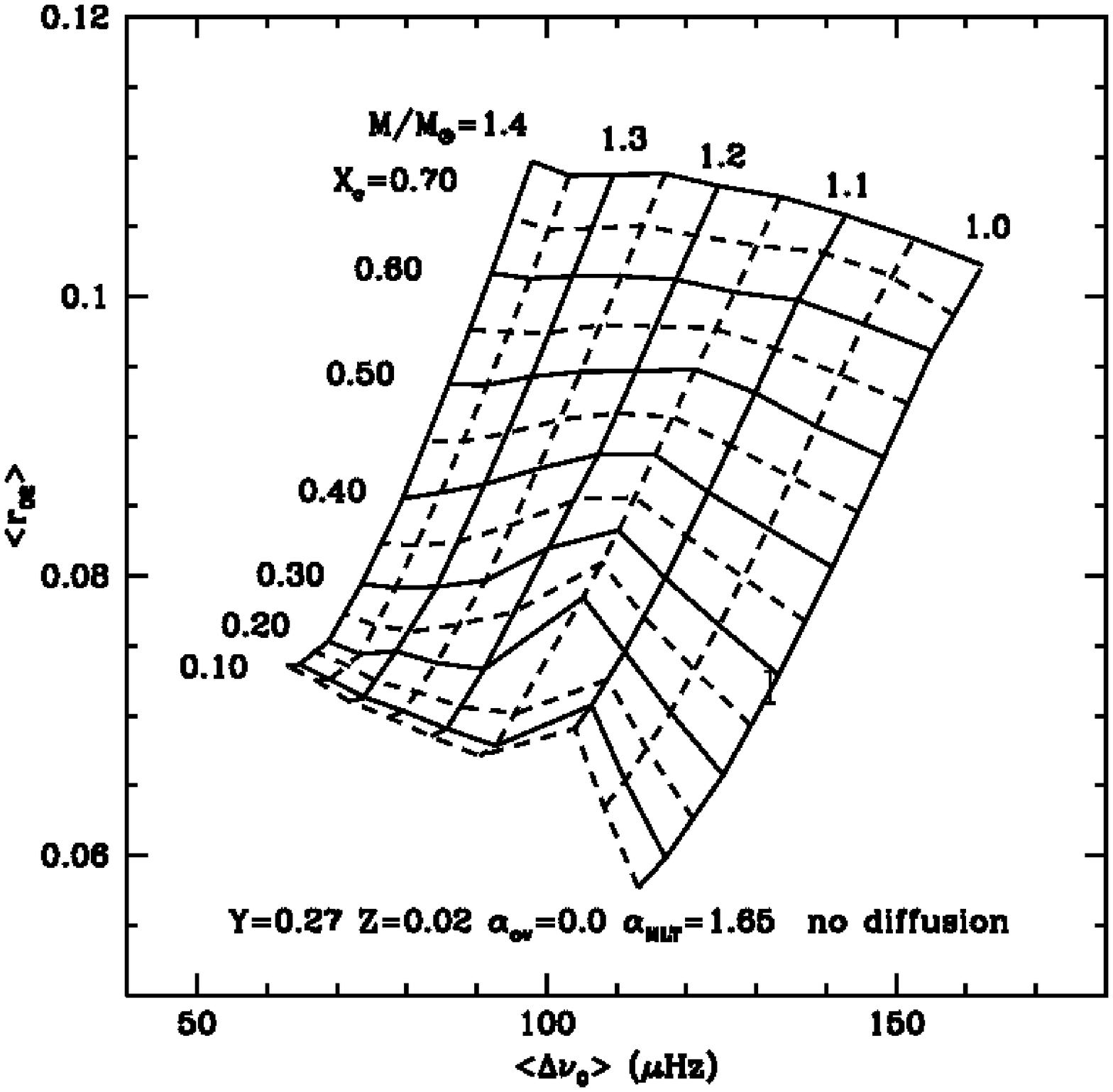}}
% \resizebox{\includegraphics[width=3.4in]{fig2a.eps} }
% \vspace*{-1.0 cm}
 \caption{Left: the inversion of isochrones in the HR diagram doesn't allow to estimate the age of non evolved stars (the age of a G-star similar to the Sun is found to be in the range 0.5-10 Gyr!) while it gives poor accuracy on the age of stars close to the turnoff (the age of a typical A-star lying at the turnoff is found to be in the range 0.6-1 Gyr). Large ellipses correspond to the present uncertainty on the HR diagram position while small ones correspond to the accuracy achievable at GAIA time (see e.g. \cite[Lebreton 2005]{2005ESASP.576..493L}). Isochrones have been obtained with the {\small CESAM} code (\cite[Morel \& Lebreton 2008]{moreleb08}). Right: Asteroseismic diagram in which the abscissa is the mean large frequency separation $\Delta\nu$ of $\ell=0$ modes of different orders and the ordinate is the mean ratio of the small separation (between $\ell=2$ and $\ell=0$ modes) to the large separation. The frequencies have been calculated for standard models of different masses and evolution stages on the MS. Models have been obtained with either the {\small CESAM} or the {\small CLES} code (\cite[Morel \& Lebreton 2008]{moreleb08}, \cite[Scuflaire et al. 2008a]{scufetal08a}) and frequencies with the {\small LOSC} oscillation code (\cite[Scuflaire et al. 2008b]{scufetal08b}). The frequency means have been estimated following \cite[Mazumdar (2005)]{Mazumdar05}. The error bar located at the solar position corresponds to the one expected from CoRoT or Kepler missions measurements, i.e. to a relative accuracy on the individual frequency of $10^{-4}$.}
   \label{fig5}
\end{center}
\end{figure}

It can be shown that the large frequency separation can be approximated by $\Delta \nu \sim (2\int_0^R \frac{dr}{c})^{-1}$. Therefore it is a measure of the inverse of the time it takes a sound wave to travel across the star and it scales as $({M}/{R^3})^{1/2}$, that is as the mean density. %$\left\langle \rho \right\rangle$. 
As shown in Fig.~\ref{fig4} left, for stars on the MS, the mean density increases as mass decreases and at given mass it decreases as evolution proceeds; we therefore expect the large separation to be quite sensitive to the mass and slightly to the evolutionary state on the MS. 
%$\langle {\rho}\rangle^{\frac{1}{2}}$
On the other hand, the small frequency separation depends on the value of $D_0= \frac{\Delta\nu}{4\pi^2\nu_{\rm n,\ell}}\left[\frac{c(R)}{R}-\int_0^R \frac{dc}{dr}\frac{dr}{r} \right]$ and is therefore sensitive to the integral over the star of the sound speed gradient weighted by $1/r$.  During MS evolution, both the temperature $T$ and the mean molecular weight $\mu$ increase in stellar cores but the relative increase in $\mu$ exceeds that in $T$ so that the net effect is a decrease of the sound speed towards the center (see Figs.~\ref{fig4} centre and right). This in turns leads to a decrease in $D_0$ and therefore of the small separation as evolution proceeds making $\delta \nu$ a valuable age estimator.

Fig.~\ref{fig5} left recalls how the age of a star is poorly constrained in those regions of the HR diagram where the isochrones are degenerate (for instance in the regions close to the zero age main sequence of low mass stars and in the turnoff regions). On the contrary if we plot in a diagram the large separation versus the small one (\cite[Christensen-Dalsgaard 1988]{JCD88}), or rather the ratio $r_{\ell, \ell+2}=\delta\nu_{\ell}/\Delta\nu_{\ell}$ of the small to the large separation which is less sensitive to surface effects (\cite[Roxburgh \& Vorontsov 2003]{RoxVor03}), we find that the degeneracy is removed (see Fig.~\ref{fig5} right). Therefore a precise measure of p-modes frequencies should provide a valuable diagnostic of the age (and mass) of solar-like oscillators. Indeed \cite[Kjedsen et al. (2008)]{2008AIPC.1043..365K} have estimated that with an accuracy on observed frequencies of a few $0.1\ \mu{\rm Hz}$ (typical accuracy of the CoRoT and future Kepler missions), the age of a solar type oscillator can be determined to better than $\approx 10\%$ of its MS lifetime.

\begin{figure}[t]
% \vspace*{-2.0 cm}
\begin{center}
\resizebox*{\hsize}{!}{\includegraphics*{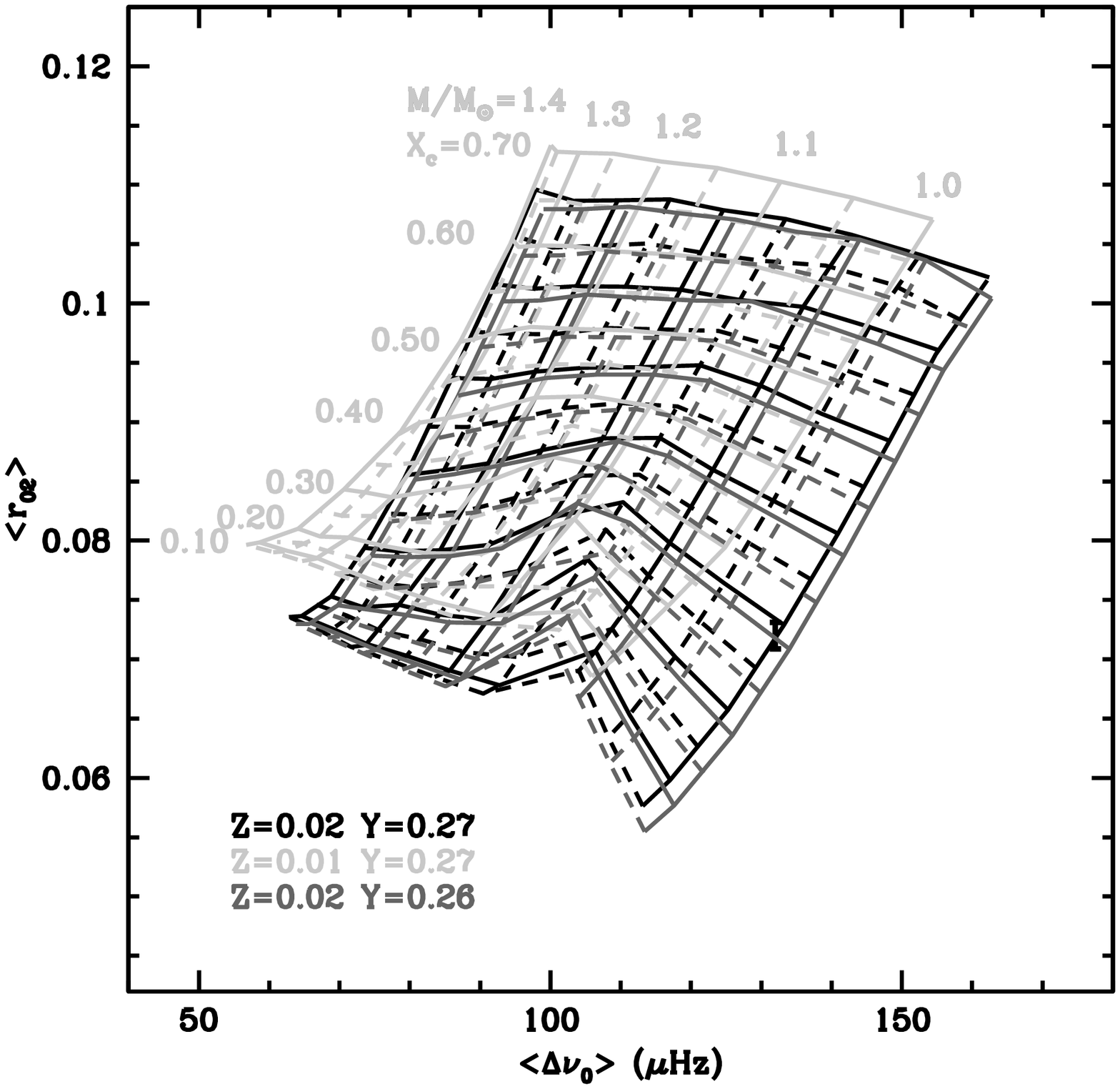}\includegraphics*{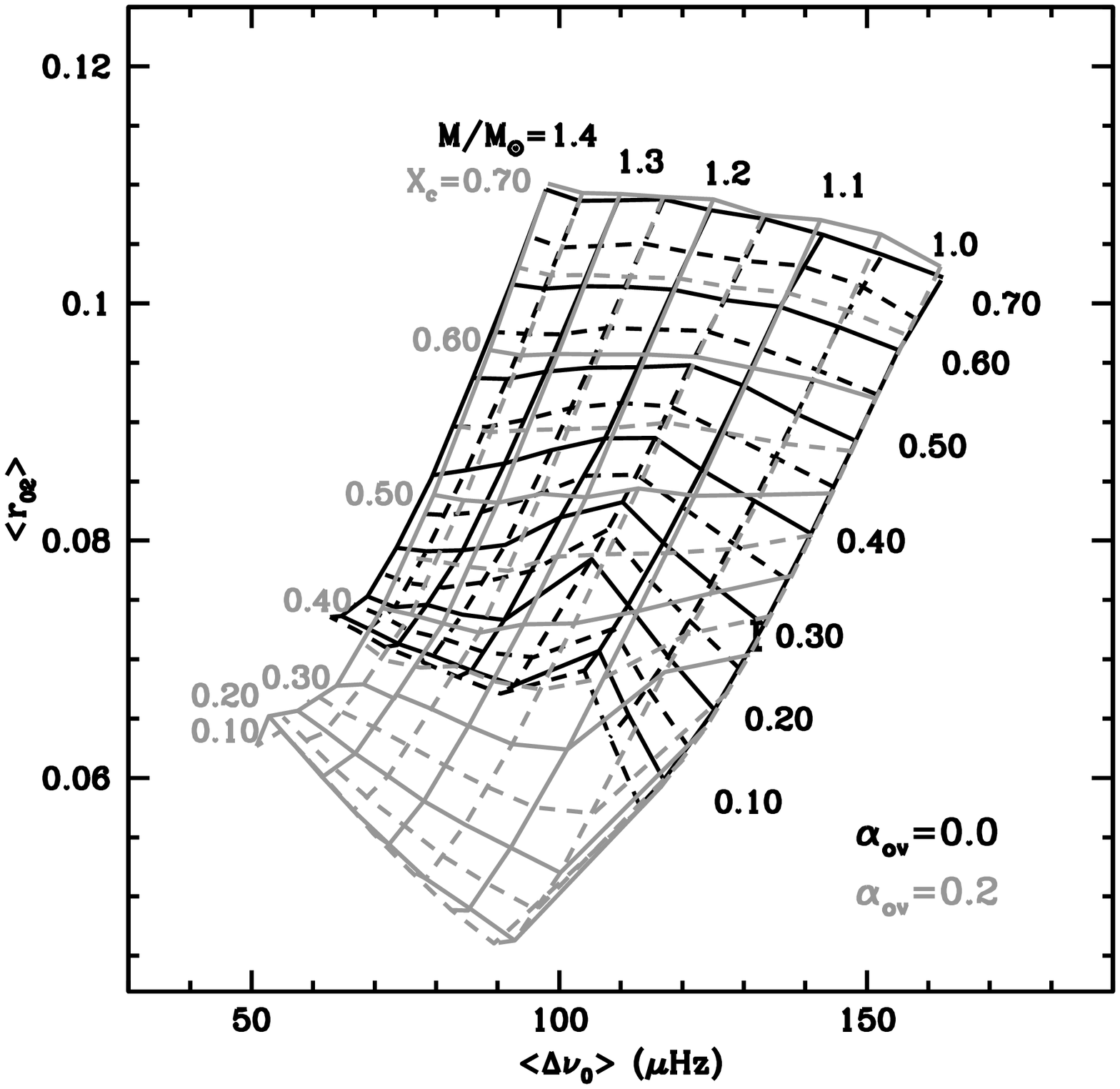}}
% \resizebox{\includegraphics[width=3.4in]{fig2a.eps} }
% \vspace*{-1.0 cm}
 \caption{Left: effects on the asteroseismic diagram of a change of chemical composition. In black models, with $Y=0.27$, $Z=0.02$; in dark grey, models with $Y=0.26$, $Z=0.02$ and in light grey, models with $Y=0.27$, $Z=0.01$. Right: Effect of overshooting with in black the models without overshooting and, in grey, models calculated with $\alpha_{\rm ov}=0.2$.}
   \label{fig6}
\end{center}
\end{figure}

In the age estimation it is important to evaluate the uncertainties that result either from the uncertainty on the determination of the stellar fundamental parameters (as the chemical composition) or from the weak knowledge of several aspects of the physics entering stellar models calculation. For instance the description of convection, the occurence of either microscopic or turbulent diffusion of chemical elements or the estimate of the size of the mixed core -which depends on processes like overshooting or rotational mixing- will all affect the calculated frequencies. We have estimated how the age estimated from an asteroseismic diagram ($\left\langle r_{0,2}\right\rangle$,$\left\langle \Delta\nu_0\right\rangle$) changes with stellar model inputs. We find that changing the mixing-length parameter from the value $\alpha_{\rm MLT}=1.65$ to $1.80$ changes the age at a given evolutionary stage on the MS (same central hydrogen content $X_{\rm c}$) by less than $1.5\%$ while introducing microscopic diffusion in the model calculation changes the age by less than $\approx5\%$. Changing the initial helium abundance in mass fraction from $Y=0.27$ to $0.26$ modifies the age at constant $X_{\rm c}$ by less than $\approx10\%$. On the other hand, as illustrated in Fig.~\ref{fig6} (left) changing the initial metallicity in mass fraction from the value $Z=0.02$ to $0.01$ changes the age by $15$ to $30\%$ while changing the overshooting parameter from $\alpha_{\rm ov}=0.0$ to $0.2$ has huge effects on the age estimate for stars possessing convective cores in the late part of their MS evolution (see Fig.~\ref{fig6} right).

To improve the determination of the age based on asteroseismic diagrams, it is therefore crucial to improve the observational determination of their metal content (this requires high resolution spectroscopy and further improvements in model atmospheres). Also it is very interesting to get additional (preferably independent) information on their fundamental parameters. This is achievable for stars observable in interferometry (direct access to the radius), for members of binary systems (access to mass and/or radius), or for hosts of exoplanets (access to the mean density from the planetary transit). Finally, as discussed in the following, the full interpretation of seismic data may allow to get further information, such as the surface helium abundance or the size of the mixed core.

\subsection{Deviations from the asymptotic theory: signature of sharp features in the interior}

Rapid variations of physical quantities in the interior of a star can be detected by seismic analysis. Indeed, any sharp feature inside the star will produce an oscillatory signal in the frequencies of the form $\widetilde{\delta \nu}^{}\sim A \cos(2\tau_{m} \omega+\phi)$ where $\tau_{m}=\int_{r_m}^R \frac{dr}{c}$ is the acoustic depth of the abrupt feature (see e.g. \cite[Gough 1990]{gough90}, \cite[Monteiro et al. 2000]{Monteiroetal00}).

For instance, the second He ionisation in the outer convection zone of solar-like stars produces a depression in the profile of the adiabatic exponent $\Gamma_1$. The higher the helium abundance in the convection zone, the deeper the $\Gamma_1$ depression. The sound speed profile locally carries the signature of the $\Gamma_1$ depression which results in an oscillatory signal in the frequencies as a function of radial order. \cite[Basu et al. (2004)]{Basuetal04} have shown that this signal, that can be extracted from low degree p modes   frequencies, can be used to infer the helium abundance in the envelope of solar-like pulsators. They find that, with a relative accuracy of $10^{-4}$ on the frequencies (typical of {CoRoT} or {Kepler} missions), it should be possible to infer the envelope helium abundance ($Y$ in mass fraction) of pulsators of mass in the range $0.8-1.4\ M_\odot$ with an accuracy of $\Delta Y\approx 0.02$ provided their radius or mass can be estimated independently. Also, it should be possible to infer the position of the basis of the convective envelope (see e.g. \cite[Ballot et al. 2004]{Ballotal04}) or the boundary of the mixed core (see e.g. \cite[Mazumdar et al. 2006]{Mazumdar06}) in a similar way. The latter depends on the extra mixing processes that can take place at the border of the convective core (overshooting, rotationally induced mixing) and it is a crucial data for age estimation.

\section{Age diagnostics through g-modes and p-g mixed modes}

The properties of g-modes are determined by the behavior of the Brunt-V\"ais\"ala frequency in the stellar interior. These modes  are sensitive probes of the chemical gradients resulting from  the combined effect of nuclear burning and  convective mixing, and hence they are  potential diagnostics of the stellar evolutionary stage (and therefore of age).   

\begin{figure}[t]
% \vspace*{-2.0 cm}
\begin{center}
\resizebox*{0.95\hsize}{!}{\includegraphics*{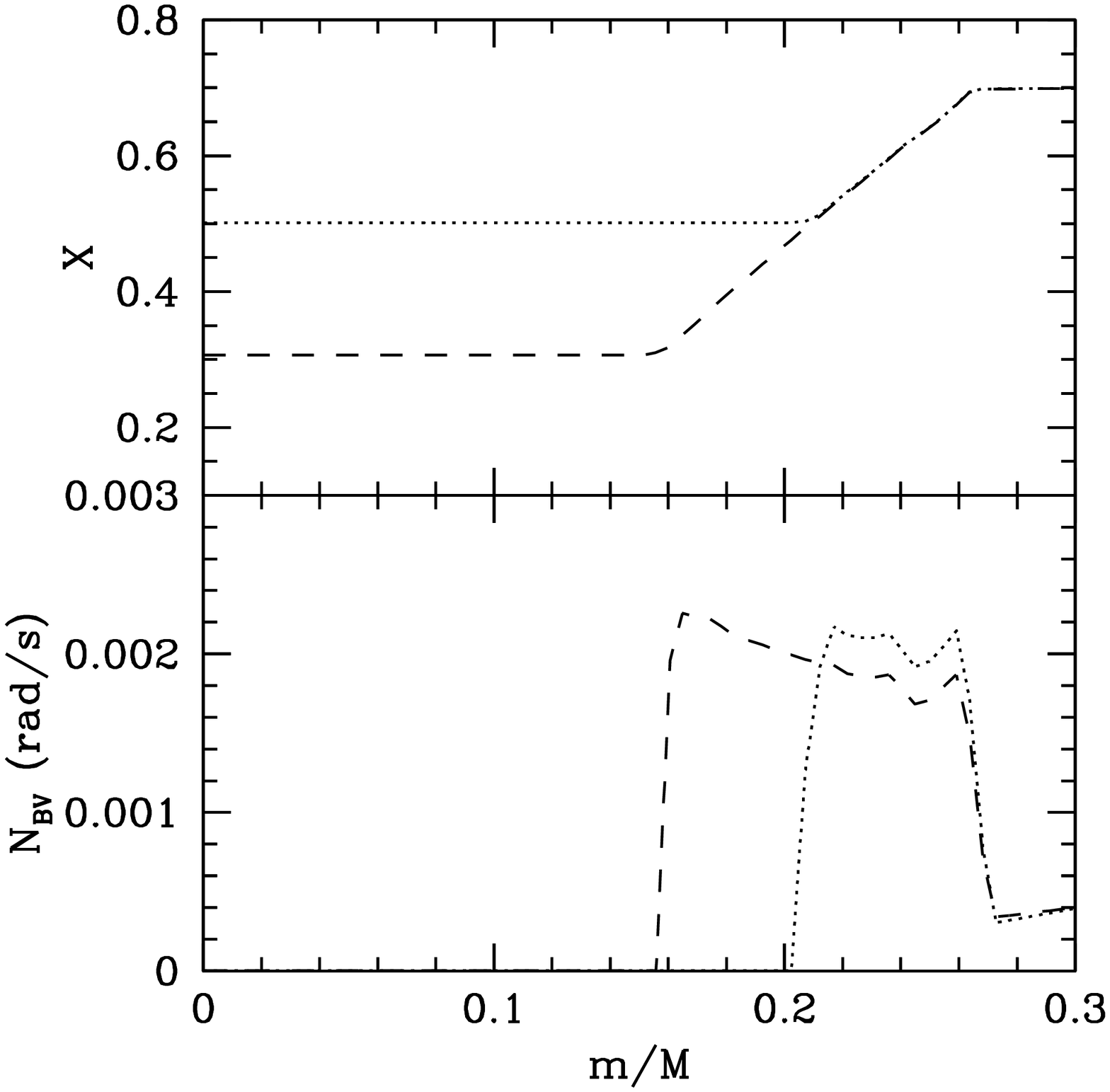}\includegraphics*{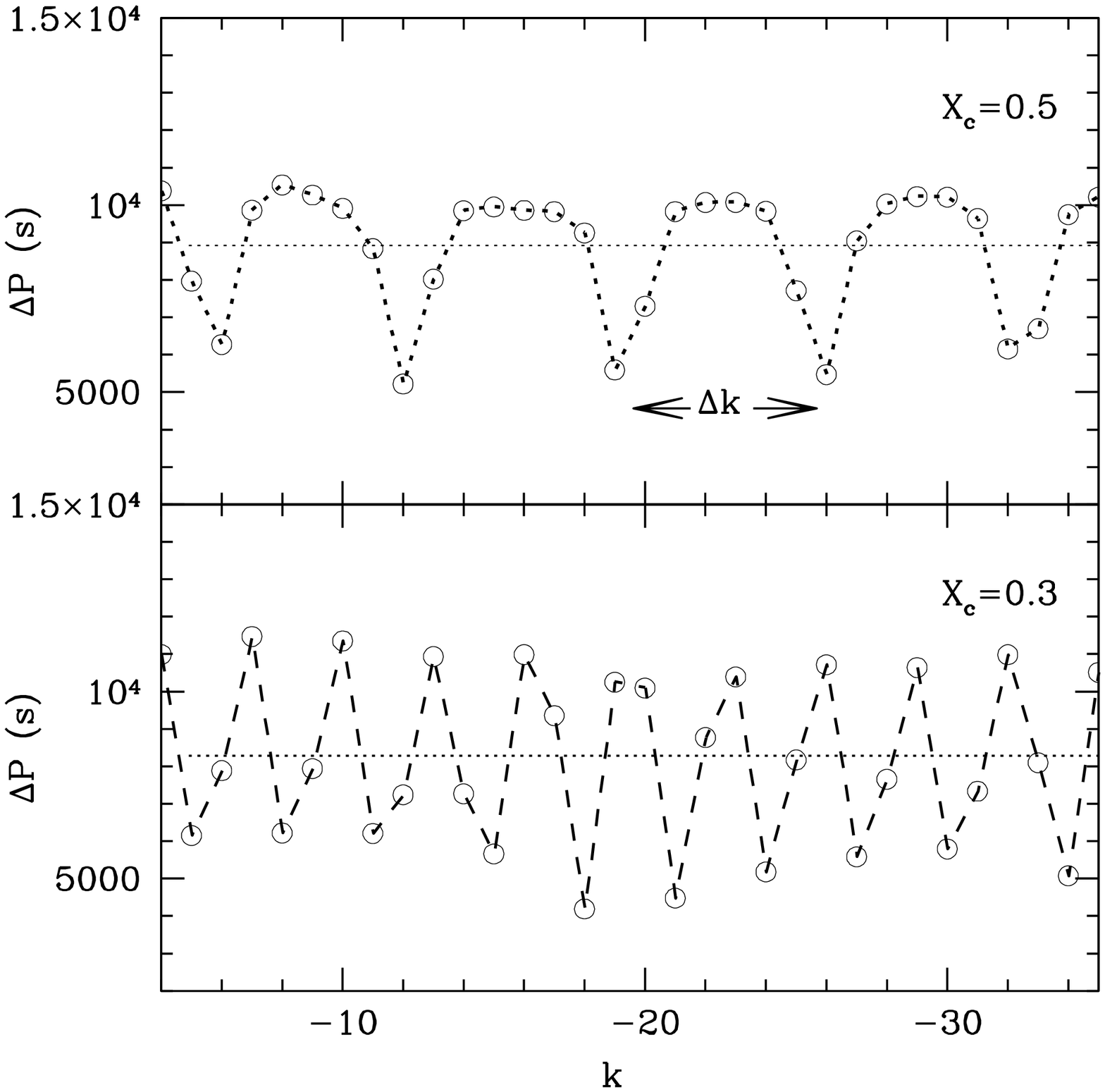}}
% \resizebox{\includegraphics[width=3.4in]{fig2a.eps} }
% \vspace*{-1.0 cm}
 \caption{Left: Hydrogen abundance in the core of 6$M_\odot$ models on the main sequence (upper panel) at $X_c\simeq 0.5$ (dotted line) and at $X_c\simeq 0.3$ (dashed line). The convective core recedes during the evolution leaving behind a chemical composition gradient and, a sharp feature in the Brunt-V\"ais\"ala frequency (lower panel). Right: Period spacing for the same models as a function of radial order $k$. Horizontal dotted lines represent constant period spacing as predicted by the asymptotic approximation, i.e. ${\Delta P_{\rm k,\ell}}\propto \Pi_0$ (\cite[Miglio et al. 2008]{Miglioetal08}).}
\label{fig7}
\end{center}
\end{figure}

\begin{figure}[t]
\vspace{-1.cm}
\parbox{7.25cm}{\includegraphics[width=7cm]{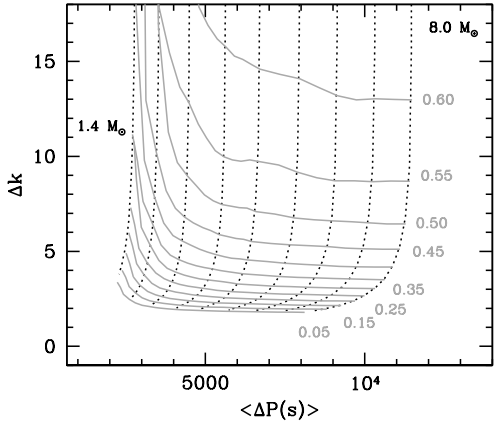}}\parbox{6.25cm}{\caption{Diagram in which the abscissa is the main period spacing for $\ell=1$, and the ordinate is the period of the oscillatory behavior of the period spacing (see Fig.~\ref{fig7}, right) in terms of the difference in radial order between two consecutive trapped modes. Dotted lines represent the variation of these quantities along the MS evolution for models with masses between 1.4 and 8~$M_\odot$,  initial chemical composition $X_0=0.70$, $Z_0=0.02$, and overshooting parameter $\alpha_{\rm OV}=0.0$. Solid lines (grey) connect models at the same evolutionary state, i.e. with the same value of the central hydrogen mass fraction.}\label{fig8}}
\end{figure}

In the case of high order low degree g-modes, such as those expected in SPB ($4-7\ M_\odot$) and $\gamma$ Doradus stars ($1.4-1.8\ M_\odot$), the first order asymptotic theory predicts that the periods of the modes should be equally spaced (\cite[Tassoul 1980]{Tassoul80}) with a period spacing  for two modes of consecutive order $k$ and $k+1$ and same degree $\ell$ given by ${\Delta P_{\rm k,\ell}}=P_{\rm k+1,\ell}-P_{\rm k,\ell}=\frac{2\pi^2}{L} \Pi_0$, with the {\it buoyancy radius} $\Pi_0=\left(\int_{r_{0}}^R \frac{\left|N_{BV}\right|}{r}dr\right)^{-1}$. This approximation is, however, no longer valid when variations of $N_{\rm BV}$ on a length scale smaller than the  oscillation wavelength are present.  That may occur in MS stars due to the building of a $\mu$-gradient at the outer border of the receding convective core. The corresponding  sharp feature in $N_{\rm BV}$ can lead to a resonant condition such that modes of different order $k$ are periodically confined in the $\nabla_{\mu}$ region. These modes have a different period and as a consequence the period spacing presents an oscillatory behavior rather than the constant value that would be expected in a model without sharp variations in $N_{\rm BV}$ (Fig.~\ref{fig7}). By using a second order approximation, it is possible to derive an analytical expression relating the amplitude and periodicity of the oscillatory component to the location and sharpness of the feature in $N_{\rm BV}$ (\cite[Miglio et al. 2008]{Miglio08} and references therein). This periodicity expressed in terms of radial order is given by the ratio between the buoyancy radius of the star, and the buoyancy radius of  the  $\nabla_{\mu}$ region: $\Delta k=\Pi_{\mu}/\Pi_0$, and corresponds to the difference between radial orders of two consecutive trapped modes in the $\nabla_{\mu}$ region. As shown in Figs.~\ref{fig7} and ~\ref{fig8}, the value of $\Delta k$ is very sensitive to the  evolutionary state of the star: for non evolved models ($X_c\sim X_0$)  $\Delta k \rightarrow \infty$ and the period spacing is almost constant as predicted by the asymptotic theory %($\left < \Delta P_{\ell}\right>$)
but when the stars evolve the periodicity of the oscillatory component in $\Delta P$ decreases. 

$\beta$~Cephei and $\delta$~Scuti show p and g-modes of low radial order and, for stars evolved enough, also p-g  mixed modes (see. Sect.~\ref{section2}). In this case, the quantitative predictions of the second order  asymptotic approximation cannot be used, but they are still able to qualitatively describe the properties of g-modes. The periods of g-modes depend on the location and shape of the chemical composition gradient and therefore so do the frequencies of mixed modes. Since the  frequency separation between consecutive modes varies rapidly during the avoided crossing phenomenon, the detection of mixed modes together with that of pure p-modes provides an important constraint to the stellar evolutionary state (see for instance \cite[Pamyatnykh et al 2004] {Pamyatnykh04}). 

Finally for low mass stars that leave the MS and are evolving as sub-giants,  the increasing central condensation together with the chemical composition gradient lead to a large increase of  $N_{\rm BV}$. As a consequence,  mixed modes  may appear in the frequency domain of solar-like oscillations. Again, the frequencies of the modes undergoing an avoided crossing  are direct probes of the evolutionary status: models on the MS do not present mixed modes and the large frequency separation is almost constant (as predicted by the asymptotic theory). On the contrary, as more evolved models are considered, the avoided crossing effects become more important and the large frequency spacing for non-radial modes becomes more irregular (see e.g. \cite[Di Mauro et al. 2004]{DiMauro04}, \cite[Miglio et al. 2007] {Miglioetal07}).

In all the cases considered above, the frequencies of  high-order g-modes and g-p mixed modes are sensitive to the location and shape of the chemical composition gradient. Therefore, these modes are sensitive probes of the evolutionary state and also of the properties of the transport processes acting in stellar cores, for instance, convective overshooting or diffusive mixing.  Nevertheless, to clearly discriminate between different scenarios, high accuracy of the classical observables such as effective temperature, luminosity and chemical composition (and  masses for binary stars) would also be required. 

Age estimates of stars in the galactic disc are crucial for galactic evolution studies.
Considering the progress expected on the determination of both the seismic and classical observables, \cite[Lebreton et al. (1995)]{lebreton95} (see also \cite[Lebreton 2005]{2005ESASP.576..493L}) find that we can reasonably expect to reduce the uncertainty on the age determination of disc A and F stars which is presently of about $35-40\%$ (of which $13-24\%$ result from the uncertainty on the size of the mixed cores) to a level of $\approx10-15\%$.

\section{Prospects and Conclusions}

We are presently in a very fruitful period where asteroseismic data are being obtained both from the ground and from space to an increasing level of precision for growing samples of stars in wide ranges of masses and evolutionary stages. The {\small{CoRoT}} satellite has observed for more than one year now. It will observe a total of more than 100 targets (solar-like stars, $\gamma$ Doradus, $\delta$ Scuti, $\beta$ Cephei, giants) in the so-called Seismo field and for stars observed during Long Runs of 150 days, a frequency accuracy as good as a few $10^{-7}$Hz will be reached (see \cite[Michel et al. 2006]{Michel06}). First {\small{CoRoT}} observations are currently under analysis and interpretation. Up to now, three solar-like pulsators have been observed (see e.g. \cite[Michel et al. 2008a]{2008Sci...322..558M}) and for the first one HD49933, observed during the first 60 days Intermediate Run and then during a 150 days Long Run, the large frequency separation has been measured and a first frequency identification for modes of degrees $\ell= 0, 1, 2$ has been proposed (\cite[Appourchaux et al. 2008]{2008A&A...488..705A}). Further very interesting results are being presently obtained related for instance to solar-like oscillations in giant F and G stars and to the very rich oscillation spectra of $\delta$ Scuti stars just revealed by {\small{CoRoT}} (see e.g. \cite[Michel et al. 2008b]{Michel08b}).

The Kepler photometry space mission to be launched in March 2009 will provide seismic data, at the same level of precision than {{CoRoT}}, for a large number of stars, hosts of exoplanets. The information on the stellar mean density, coming from the observation of the exoplanet transit, will add constraints for the age determination. Asteroseismic observations are presently obtained in spectroscopy with {\small{HARPS, CORALIE, ELODIE, UVES, UCLES}} and in photometry with {\small{MOST, CoRoT, WIRE}} while other projects are about to begin or are  currently under study either in spectroscopy ({\small{SONG, SIAMOIS}}) or in photometry ({\small{Kepler, PLATO}}).

In parallel, much progress will be done in the determination of the stellar global parameters. For instance GAIA (ESA 2000; \cite[Perryman et al. 2001]{Perry01}), to be launched in 2011, will make astrometric measurements, at the micro-arc second level together with photometric and spectroscopic observations of a huge number of stars covering the whole range of stellar masses, compositions and evolution stages. Important progress is also expected from interferometric and high resolution spectroscopy measurements (for instance with {\small{VLT-VLTI, CHARA, KECK, JWST}} etc.).

Asteroseismology provides diagnostics of stellar evolution all across the HR diagram. If we consider both the present and expected future accuracies on the observed frequencies \underline{and} the future improvements expected in the determination of the stellar global parameters, it seems reasonable to forecast that asteroseismology will allow to determine the ages of individual stars with a $10-20\%$ accuracy. This will certainly lead to significant progress in the understanding of the history and evolution of galaxies.

\begin{discussion}
\end{discussion}

\end{document}